\documentclass[journal]{ieeeconf}  

\IEEEoverridecommandlockouts                              
\usepackage{graphics} 
\usepackage{epsfig} 
\usepackage{mathptmx} 
\usepackage{times} 
\usepackage{amsmath} 
\usepackage{amssymb}  
\usepackage{color}
\usepackage{multirow}
\usepackage{footnote}
\usepackage{caption}
\usepackage[hyphens]{url}
\usepackage{hyperref}
\usepackage{float}
\usepackage{cuted}
\usepackage[table,xcdraw]{xcolor}
\usepackage{subcaption}
\usepackage{comment}
\usepackage{cite}
\usepackage{soul}

\definecolor{boristext}{rgb}{0.22, 0.44, 0.88}
\definecolor{boriscomments}{rgb}{0.88, 0.04, 0.04}
\definecolor{borishighlights}{rgb}{0.2, 0.88, 0.2}
\newcommand{\bb}[1]{\textcolor{boristext}{#1}} 
\newcommand{\bcom}[1]{\textcolor{boriscomments}{[#1]}} 

\usepackage[export]{adjustbox}

\title{\LARGE \bf
Time-Sensitive Networking in IEEE 802.11be: \\ On the Way to Low-latency WiFi~7}

\author{Toni Adame, Marc Carrascosa, and Boris Bellalta\\Department of Information and Communication Technologies, Universitat Pompeu Fabra\\Email: \{toni.adame, marc.carrascosa, boris.bellalta\}@upf.edu}

\begin{document}

\maketitle
\thispagestyle{empty}
\pagestyle{empty}

\begin{abstract}
Short time after the official launch of WiFi 6, IEEE 802.11 working groups are already designing its successor in the wireless local area network (WLAN) ecosystem: WiFi~7. With the IEEE 802.11be amendment as one of its main constituent parts, future WiFi~7 aims to include time-sensitive networking (TSN) capabilities to support low latency and ultra reliability in license-exempt spectrum bands. This article first introduces the key features of IEEE 802.11be, which are then used as the basis to discuss how TSN functionalities could be implemented in WiFi~7. Finally, benefits and requirements of the most representative low-latency use cases for WiFi~7 are reviewed. 

\end{abstract}

\section{Introduction}


In recent years, an increasing number of heterogeneous productive sectors, motivated by the latest technological advances on multimedia, cloud computing, artificial intelligence, automation, robotics, and unmanned vehicles, are fostering the emergence of cutting-edge real-time applications which strongly depend on extremely low latency and, occasionally, very high bandwidth-demanding communications for their successful operation.

Since its emergence in the early 2000s, WiFi worldwide success has been mainly substantiated on its high flexibility, mobility of devices, better cost efficiency, and reduced complexity. Although WiFi has been constantly evolving through successive amendments to improve peak throughput, capacity, and efficiency, it has not yet been able to produce a similar solution to manage time-sensitive traffic with bounded low latency.

To address the requirements of emerging real-time applications within IEEE 802.11-based networks, initiatives like the Real Time Application Technical Interest Group (RTA TIG)~\cite{meng2019ieee} are promoting physical (PHY) and medium access control (MAC) enhancements, as well as new capabilities under the time-sensitive networking (TSN) framework. Originally intended for Ethernet, TSN sub-standards, which ensure zero packet loss due to buffer congestion, extremely low packet loss due to equipment failure, and guaranteed upper bounds on end-to-end latency~\cite{finn2018introduction}, are now making their way to wireless networks.

IEEE P802.11be Task Group (TGbe)~\cite{IEEE2019TGbe} was created in May 2019 to address the design of a new PHY and MAC amendment. Considered as the successor of IEEE 802.11ax~\cite{bellalta2016ieee} and the core piece of next WiFi~7, IEEE 802.11be aspires to achieve a peak throughput of 30~Gbps and incorporate disruptive solutions in the WiFi ecosystem such as multi-link operation and multi-access point (multi-AP) coordination. At the same time, IEEE 802.11be also targets reducing worst-case latency and jitter in wireless local area networks (WLANs), for which TSN sub-standards are currently under study for their possible adoption.

Indeed, to be used as part of a potential IEEE 802.11be low-latency operation mode, original TSN mechanisms will need to be redesigned taking into consideration the inherent constraints of the wireless medium (namely, unreliability of links, asymmetric path delay, channel interference, signal distortion, lack of accurate clock synchronization methods, and incompatibility of network interface cards)~\cite{mildner2019time}, while ensuring backward compatibility with legacy WiFi devices.



Overall, wireless TSN opens new research directions for the upcoming years, and not only in the WiFi ecosystem. In fact, the 3GPP mobile standards body has also defined ultra-reliable low-latency communications (URLLC) as one of the main application areas for the enhanced capabilities of 5G. Latency reduction techniques and support to deterministic communications are also since long ago in the spotlight of low-power wireless sensor networks, particularly as a result of the specialized MAC-layer profiles introduced in IEEE 802.15.4e.





This paper introduces the future IEEE 802.11be amendment, discussing how its new features can be used to support a seamless adoption of TSN mechanisms. Hence, whereas WiFi networks will never be able to offer bounded delay guarantees due to their own nature and operation in license-exempt bands, the adoption and integration of TSN concepts would keep WiFi as one of the leading wireless access technology in the 6G era. 

The remainder of this article is organized as follows: Section~\ref{limitations} overviews the limitations of current WLANs to handle time-sensitive traffic. Section~\ref{11be} describes the main features of IEEE 802.11be in terms of PHY and MAC layers. A brief description of TSN and the potential enhancements to support it in WiFi~7 are provided in Section~\ref{supportingTSN}. The most representative WiFi~7 use cases that could leverage low-latency communications are reviewed in Section~\ref{usecases}. Lastly, Section~\ref{conclusions} presents the obtained conclusions and discusses open challenges.


\section{Limitations of IEEE 802.11 to handle time-sensitive traffic}
\label{limitations}

IEEE 802.11 offers great accessibility and ease of use, creating an open environment for any station (STA) willing to associate to the network. But at the same time, the wireless medium is precisely the main cause that hinders proper delivery of time-sensitive traffic, due to its variable capacity (which depends on the link quality) and typically higher PER (due to the stochastic properties of the channel and the presence of interference)~\cite{cavalcanti2019extending}.

As for the MAC layer, IEEE 802.11 has traditionally relied on the distributed coordination function (DCF): a contention-based random access scheme based on carrier sense and exponential backoff rules. The main drawback of DCF, however, is its non-predictable behavior and lack of traffic prioritization techniques. In fact, in presence of multiple STAs, DCF may lead to channel saturation by contending packets, thus being unable to guarantee timely data delivery. The alternative point coordination function (PCF), based on a centralized polling system, has never been widely adopted.

The enhanced distributed channel access (EDCA) was envisioned as part of the IEEE 802.11e amendment to extend DCF and provide quality of service support according to 4 differentiated access categories (ACs): background, best effort, video, and voice. Prioritization is then implemented by allocating different contention-related parameters to each AC. Nevertheless, the low number of ACs, the lack of mechanisms for the prioritization of different streams belonging to the same AC, and (in some hardware devices) the use of a single buffer to store packets with different priorities are among the main EDCA shortcomings.


To outperform IEEE 802.11e operation for real-time multimedia content delivery, IEEE 802.11aa introduced the intra-AC traffic differentiation functionality, with the definition of two new time-critical voice and video ACs. However, in general, none of IEEE 802.11 mechanisms guarantee the quality of service of heterogeneous real-time streams when a WLAN is overloaded~\cite{costa2015limitations}. In such cases, flexible scheduling policies and/or admission control algorithms are highly required to effectively manage different traffic flows. 

Neighboring networks represent a key limitation to provide low-latency guarantees in all the aforementioned channel access methods. In dense scenarios, overlapping of basic service set (BSS) coverage areas turns into large delays for STAs waiting to access the channel. IEEE 802.11ax partially addresses this issue by allowing concurrent transmissions under the spatial reuse scope, showing a clear gain for time-sensitive communication~\cite{wilhelmi2019spatial}. A gain that could be remarkably boosted by means of coordination mechanisms among neighboring APs.

And lastly, when it comes to the transport layer, the \textit{bufferbloat} problem may prevent IEEE 802.11 networks from delivering time-sensitive traffic in presence of TCP flows, due to the high latency produced by excessive buffering of packets. In fact, well-known techniques to mitigate this problem in wired networks (e.g., decreasing buffer sizes and/or applying modern queue management algorithms) have proven low success in WiFi~\cite{hoiland2017ending}. 



\section{IEEE 802.11be}
\label{11be}

This section introduces the main technologies under discussion in TGbe for both PHY and MAC layers and discusses to what extent they would help to satisfy low-latency requirements. In general terms, and following the traditional IEEE 802.11 evolution, IEEE 802.11be will adopt IEEE 802.11ax contributions, further refining and extending them, and adding some new features~\cite{khorov2020current}.

\subsection{PHY layer}





The ongoing release of the 6~GHz band throughout the world will be of great benefit to WiFi dense scenarios, not only due to the additional 1.2~GHz of available spectrum, but also to the resulting interference reduction among networks/BSSs. The incorporation of the 6~GHz band into IEEE 802.11be will also encompass channels as wide as 320~MHz, hence enabling higher transmission rates. 





As for the maximum number of spatial streams, it is expected to double its number from 8 in IEEE 802.11ac/ax to 16 in IEEE 802.11be, thus further benefiting from fundamental advantages of predominantly indoor WiFi operation: rich scattering, higher angular spreads, lower correlation, and diversity of channels with good propagation conditions. 

The maximum supported modulation size in IEEE 802.11be is likewise expected to be boosted with the adoption of the 4096-QAM modulation, whose practical use, however, will only be feasible in combination with beamforming.


All in all, new IEEE 802.11be PHY features favor low-latency operation, as (1) wider available bandwidth results in faster transmissions and (2) more spatial streams turn into higher rates in the single-user (SU) mode and into more parallel transmissions (with less waiting time in the buffer) in the multi-user (MU) mode.


    %

\subsection{MAC layer}


Many significant MAC features from IEEE 802.11ax such as MU-MIMO, OFDMA, and spatial reuse will be extended in IEEE 802.11be. The support of more spatial streams will also enable more flexible MU-MIMO arrangements. However, current explicit channel state information acquisition procedure may not cope well with such high number of antennas and, for that reason, TGbe is currently evaluating several alternatives to enhance explicit sounding, even considering the introduction of an implicit procedure. 

As for OFDMA, enhanced resource unit (RU) allocation schemes will allow to allocate multiple contiguous and non-contiguous RUs to a single STA. Consequently, these novel schemes could significantly increase spectral efficiency and overall network throughput, and even better satisfy timely data delivery~\cite{avdotin2019enabling}. In fact, whether based on MU-MIMO or OFDMA, MU transmissions are key to reduce the channel access latency, as packets from different users can be dequeued simultaneously. 

\begin{figure*}[h]
    \centering
    \includegraphics[width=0.9\textwidth]{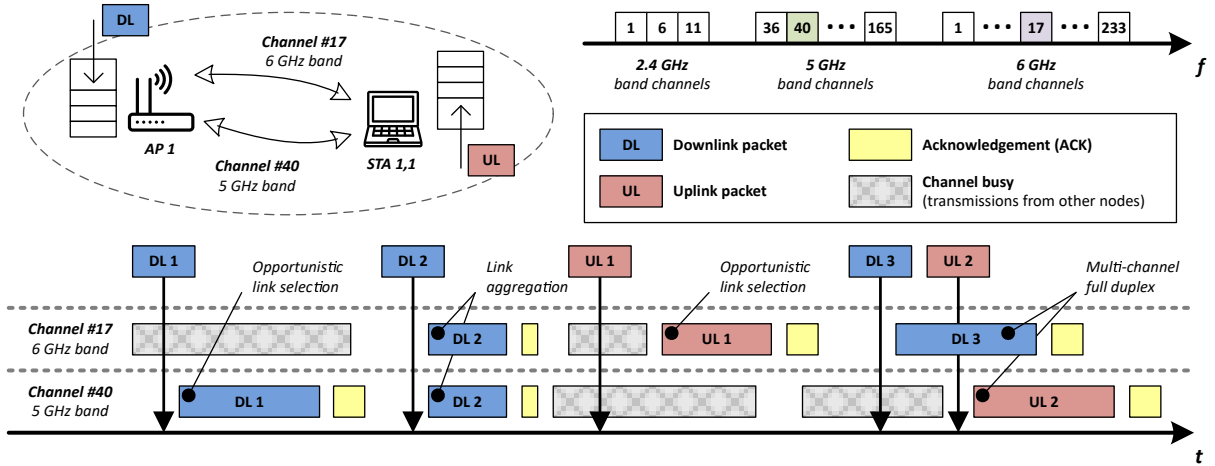}
    \caption{Multi-link operation techniques at ISM frequency bands.}
    \label{fig:multi}
\end{figure*}

Multi-link operation will likely become the most representative feature of IEEE 802.11be, aiming to 1) improve throughput by aggregating links, 2) enhance reliability by transmitting multiple copies of the same frame in separated links, and 3) decrease channel access delay by selecting the first available link in terms of latency~\cite{lopez2019ieee}. As it can be seen in Figure~\ref{fig:multi}, having two active links operating at different bands between an AP and an STA may increase channel access efficiency by enabling opportunistic link selection, link aggregation, and multi-channel full duplex.


TGbe also considers multi-AP coordination, which consists in the cooperative use of neighboring APs in enterprise IEEE 802.11be WLANs, as a way to improve overall performance by means of different techniques:

\begin{itemize}
    \item \textit{Coordinated spatial reuse (CSR)} consists in jointly negotiating the transmission power of potential overlapping APs to reduce overall interference. Access delay to the medium could be then reduced, since CSR allows to increase the number of concurrent transmissions.
    \item \textit{Coordinated OFDMA (Co-OFDMA)} optimizes the efficiency of the wireless spectrum both in time and frequency, as APs are able to allocate the available RUs to their corresponding STAs in a coordinated way. In consequence, time-sensitive and best-effort traffic could be provided with differentiated RUs to meet timely delivery requirements.
    \item \textit{Coordinated beamforming (CBF)} enables simultaneous transmissions within the same coverage area while ensuring spatial radiation nulls to non-targeted devices.
    \item \textit{Distributed MU-MIMO} allows APs to perform joint data transmissions to multiple STAs by reusing the same time/frequency resources. Spatial diversity can then be exploited to increase frame reception probability. 
\end{itemize}

Thanks to the multi-AP coordination, multiple overlapping BSSs (OBSSs) can turn channel contention in our favor, resulting in a better use of shared resources. Beyond the latency reduction obtained by using the spectrum more efficiently, new solutions to protect time-critical traffic across the cooperating BSSs may be enabled. For instance, APs dealing with best-effort traffic may agree on reducing transmission power to provide spatial reuse opportunities, so that STAs from other BSSs can successfully transmit their short-duration, time-sensitive packets at the same time.


Advanced transmission schemes such as hybrid automatic repeat request (HARQ) offer notable performance gains in varying channels compared to the traditional \textit{stop \& wait} approach, but it is not yet clear if such gains will also be achieved in WLANs due to the severity of collisions. Be that as it may, HARQ still retains the prospect of improving performance in terms of latency, because of its ability to reduce the number of required retransmissions per packet. 




\begin{table*}[h]
\scriptsize
\centering
\caption{Potential enhancements to support TSN in WiFi~7.}
\label{tab:enhancements}
\begin{tabular}{|l|l|l|c|c|c|c|}
\hline
\rowcolor[HTML]{C0C0C0} 
\cellcolor[HTML]{C0C0C0}                                                                                                          & \cellcolor[HTML]{C0C0C0}                                                                                          & \cellcolor[HTML]{C0C0C0}                                                & \multicolumn{4}{c|}{\cellcolor[HTML]{C0C0C0}\textbf{Targeted feature}}          \\ \cline{4-7} 
\rowcolor[HTML]{C0C0C0} 
\multirow{-2}{*}{\cellcolor[HTML]{C0C0C0}\textbf{Component}}                                                                      & \multirow{-2}{*}{\cellcolor[HTML]{C0C0C0}\textbf{Subcomponent}}                                                   & \multirow{-2}{*}{\cellcolor[HTML]{C0C0C0}\textbf{Potential enhancement}} & \textbf{Latency} & \textbf{Jitter} & \textbf{Reliability} & \begin{tabular}[c]{@{}c@{}}\textbf{Resource} \\ \textbf{management}\end{tabular} \\ \hline
\multicolumn{2}{|l|}{\cellcolor[HTML]{EFEFEF}}                                                                                                                                                                                                        & IEEE 802.1AS over IEEE 802.11                                 & x                & x               &                      &                     \\ \cline{3-7} 
\multicolumn{2}{|l|}{\multirow{-2}{*}{\cellcolor[HTML]{EFEFEF}\textbf{Time synchronization}}}                                                                                                                                                         & IEEE 802.11mc FTM                                             & x                & x               &                      &                     \\ \hline
\cellcolor[HTML]{EFEFEF}                                                                                                          & \cellcolor[HTML]{EFEFEF}\textbf{\begin{tabular}[c]{@{}l@{}}Traffic \\ prioritization \end{tabular}}                & EDCA operation enhancements                                    & x                &                 &                      &                     \\  \cline{2-7}
\cellcolor[HTML]{EFEFEF}                                                                                                          & \cellcolor[HTML]{EFEFEF}\textbf{\begin{tabular}[c]{@{}l@{}}Frame \\ preemption\end{tabular}}                     & IEEE 802.1Qbu-based frame preemption                                               & x                & x               &                      &                     \\ \cline{2-7} 
\cellcolor[HTML]{EFEFEF}                                                                                                          & \cellcolor[HTML]{EFEFEF}                                                                                          & IEEE 802.11e/aa admission control                            & x                &                 & x                    &                     \\ \cline{3-7} 
\cellcolor[HTML]{EFEFEF}                                                                                                          & \multirow{-2}{*}{\cellcolor[HTML]{EFEFEF}\textbf{\begin{tabular}[c]{@{}l@{}}Admission\\ control\end{tabular}}}    & Multi-band admission control                                   & x                &                 & x                    &                     \\ \cline{2-7} 
\cellcolor[HTML]{EFEFEF}                                                                                                          & \cellcolor[HTML]{EFEFEF}                                                                                          & IEEE 802.1Qbv-based time-aware shaper                                         & x                & x               & x                    &                     \\ \cline{3-7} 
\cellcolor[HTML]{EFEFEF}                                                                                                          & \cellcolor[HTML]{EFEFEF}                                                                                          & Trigger-based access                                           & x                & x               & x                    &                     \\ \cline{3-7} 
\multirow{-8}{*}{\cellcolor[HTML]{EFEFEF}\textbf{\begin{tabular}[c]{@{}l@{}}Traffic \\ shaping\\ and \\ scheduling\end{tabular}}} & \multirow{-3}{*}{\cellcolor[HTML]{EFEFEF}\textbf{\begin{tabular}[c]{@{}l@{}}Scheduled \\ operation\end{tabular}}}   & TWT mechanism                                                  & x                & x               & x                    &                     \\ \hline 

\multicolumn{2}{|l|}{\cellcolor[HTML]{EFEFEF}}                                                                                                                                                                                                        & Multi-link operation                       &     x             &                 & x                    &                     \\  \cline{3-7} 
\multicolumn{2}{|l|}{\cellcolor[HTML]{EFEFEF}}                                                                                                                                                                                                        & HARQ                                   &      x              &                & x                    &                       \\ \cline{3-7} 
\multicolumn{2}{|l|}{\cellcolor[HTML]{EFEFEF}}                                                                                                                                                                                                        & Rate adaptation in trigger-based access                                  &                 &                & x                    &                       \\ \cline{3-7} 

\multicolumn{2}{|l|}{\multirow{-4}{*}{\cellcolor[HTML]{EFEFEF}\textbf{Ultra reliability}}}                                                                                                                                                            & \multirow{-3}{*}{}IEEE 802.1CB adaptations                        &                  &                 & x                    &                     \\ \hline

\cellcolor[HTML]{EFEFEF}                                                                                                          & \cellcolor[HTML]{EFEFEF}                                                                                          &  CSR                                           &     x           &     x          &               &     x                \\ \cline{3-7} 
\cellcolor[HTML]{EFEFEF}                                                                                                          & \cellcolor[HTML]{EFEFEF}                                                                                          & Co-OFDMA                                             &      x         &    x           &     x              &           x          \\ \cline{3-7} 
\cellcolor[HTML]{EFEFEF}                                                                                                          & \cellcolor[HTML]{EFEFEF}                                                                                          & CBF                                              &      x           &                &        x             &              x    \\ \cline{3-7} 
\multirow{-4}{*}{\cellcolor[HTML]{EFEFEF}\textbf{\begin{tabular}[c]{@{}l@{}}Resource \\ management\end{tabular}}}                & \multirow{-4}{*}{\cellcolor[HTML]{EFEFEF}\textbf{\begin{tabular}[c]{@{}l@{}}Multi-AP \\ resource \\ coordination\end{tabular}}}                             &          Distributed MU-MIMO      &            x     &                      &       x      &  x     \\ \hline
\end{tabular}
\end{table*}


In short, the new IEEE 802.11be MAC functionalities will help to use more efficiently the spectrum resources and allocate them in a more flexible way to optimize throughput, latency, or reliability, depending on the scenario requirements. Furthermore, these functionalities can be better exploited for low-latency purposes if some core TSN features (e.g., admission control and scheduled operation) are integrated on top of them, as we will see in Section~\ref{supportingTSN}.


Last but not least, the lack of legacy devices operating in the upcoming 6~GHz band also offers the possibility of rethinking channel access for future WiFi~7 adopters. In this sense, traditional channel access schemes based on contention might be partially replaced by others able to offer higher levels of determinism, thus facilitating the management of real-time deterministic traffic and the inclusion of TSN mechanisms.

\subsection{Standardization status}

The standardization process of IEEE 802.11be, initiated by TGbe in May 2019, consists of two stages: Release~1 and 2, and it is expected to be completed in May 2024 with the publication of the final amendment. Release~1 is aimed to prioritize the development of a small distinctive set of IEEE 802.11be candidate features, such as the 320~MHz channels, the 4096-QAM modulation, and the multi-link operation, becoming available by 2022. Release~2 shall contain the rest of the features (including a low-latency operation mode) as well as the potential extensions and/or modifications of the already introduced ones in Release~1.

\section{Supporting TSN in WiFi~7}
\label{supportingTSN}

TSN consists of a set of sub-standards defined by the IEEE 802.1 TSN Task Group~\cite{IEEE2019TSN} to support deterministic messaging on standard Ethernet. Essentially, TSN technology relies on a central management that uses time scheduling to ensure reliable packet delivery with bounded latency and low packet delay variation (\textit{jitter}) in deterministic real-time applications.\footnote{Ultra reliability in TSN is responsibility of IEEE 802.1CB, which basically sends duplicate copies of each frame over disjoint wired paths to provide proactive seamless redundancy.} The coexistence of different traffic classes is guaranteed by two TSN sub-standards: whereas IEEE 802.1Qbu implements frame preemption to interrupt any ongoing operation if a time-sensitive frame is selected for transmission, IEEE 802.1Qbv creates exclusive time slots for time-sensitive frames managed by a time-aware shaper.

A careful design of WiFi~7 technologies taking into account the TSN principles could certainly contribute to reduce WiFi latency issues, yet at the present time that potential integration is neither straightforward nor exempted from uncertainties and incompatibilities. The approach that could be followed by TGbe encompasses both adaptations from TSN sub-standards and proposals of new solutions in several areas, as compiled in Table~\ref{tab:enhancements}.

This section elaborates on the time synchronization and the traffic shaping and scheduling components of TSN, analyzes the most suitable IEEE 802.11be enhancements to support them, points out the main challenges involved in the integration process, and sheds some light on possible solutions and open research directions.



\subsection{Time synchronization}

TSN sub-standard IEEE 802.1AS includes a version of the precision time protocol (PTP), which enables the distribution of a single reference clock across network devices in a master/slave basis. The availability of a common clock is likewise a key requirement for WiFi~7, as it would permit to successfully schedule MU transmissions in both uplink and downlink, as well as to establish coordination mechanisms among APs.

Indeed, IEEE 802.1AS can already be operated over IEEE 802.11 by means of the timing measurement (TM) procedure defined in IEEE 802.11v, which takes wireless link asymmetric delay into consideration. Time is propagated in private action frames between a master (i.e., the AP) and a slave (i.e., the STA), being the latter able to compute the clock offset and adjust its own time accordingly.

Furthermore, the next revision of the IEEE 802.1AS standard (IEEE 802.1AS-Rev, still in draft version) will contain a novel synchronization method by using the IEEE 802.11mc fine timing measurement (FTM) procedure. FTM provides $0.1$~ns of timestamp resolution, far more accurate than TM, whose timestamp resolution is 10~ns~\cite{mahmood2016clock}.

\subsection{Traffic prioritization}

The four access categories (ACs) employed by EDCA are insufficient for fine control of real-time applications, as they cannot provide hard bounds on latency/jitter, especially under congestion. For that reason, TGbe is considering possible enhancements to EDCA, such as the incorporation of a new AC with the highest priority for time-sensitive traffic~\cite{genc2019wi}, as well as some modifications to how obtained transmission opportunities (TXOPs) are used (for instance, to allow using any TXOP, regardless the AC that has obtained it, to send time-sensitive traffic when available~\cite{khorov2020current}).

\begin{figure*}[h]
        \centering 
        \includegraphics[width=0.7\textwidth]{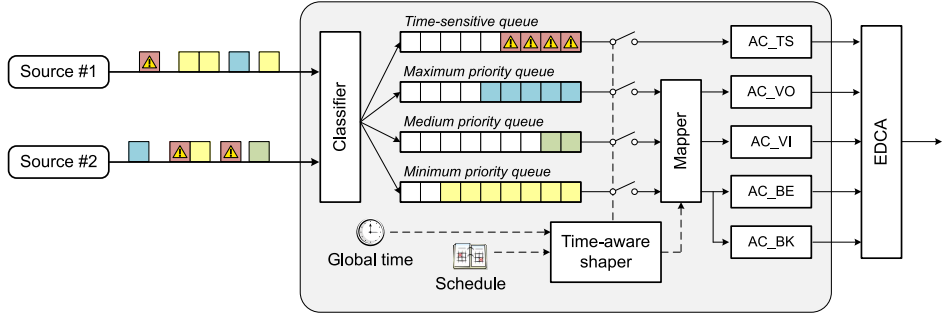}
    \caption{TSN-based traffic classification and scheduling over EDCA.}
        \label{fig:scheduling}
\end{figure*}

In addition, the adaptation of the IEEE 802.1Qbv time-aware shaper on top of one of the IEEE 802.11 MAC modes would allow devices to control how traffic arrives to the different EDCA ACs according to new rules yet to be defined. By shaping the traffic that arrives to the MAC layer, devices will be able to reduce inter-AC contention, as well as better control how and when they content for the channel. On this basis, Figure~\ref{fig:scheduling} exemplifies the hypothetical integration and joint operation of a TSN-based time-aware shaper with EDCA.





\subsection{Frame preemption}

\begin{figure}[]{}
    \centering
    \includegraphics[width=0.45\textwidth]{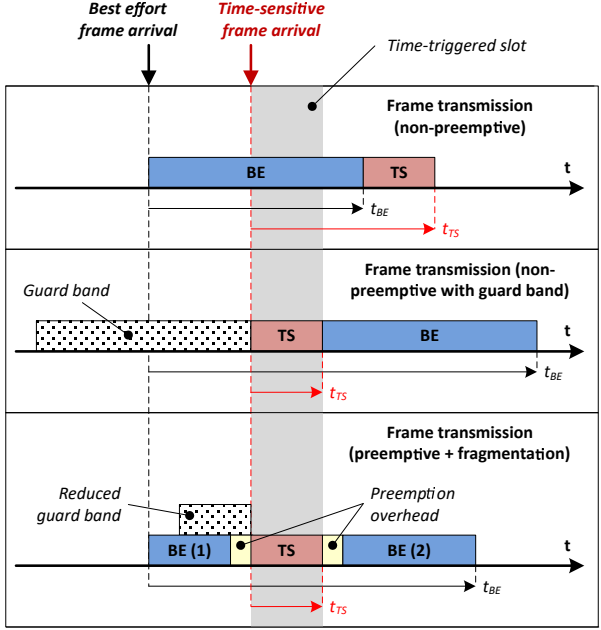}
    \caption{Frame preemption mechanism, being \textit{BE}: Best-effort frame, \textit{TS}: Time-sensitive frame, $t_{\text{BE}}$: Best-effort frame latency, and $t_{\text{TS}}$: Time-sensitive frame latency.}
    \label{fig:preemption}
\end{figure}

In the case a node is transmitting multiple traffic flows, placing the time-sensitive traffic in the highest priority queue may not be enough to mitigate the residual delay caused by large ongoing low-priority transmissions, which may include many aggregated packets and last up to the maximum physical protocol data unit (PPDU) duration (i.e., $\sim 5$~ms). As shown in Figure~\ref{fig:preemption}, a possible solution to that issue could be based on the adaptation of the IEEE 802.1Qbu frame preemption mechanism, which would also foster the use of packet aggregation even in presence of time-sensitive traffic, thus improving overall throughput.

Integrating frame preemption into WiFi would require, however, several changes in the physical and link layers, such as the format of preemptable frames and the methods to fragment frames while preserving integrity of preemptable traffic. In any case, it seems reasonable to only support this new feature when aggregate MAC protocol data units (A-MPDUs) are transmitted, and so extend the service field used to identify the different MPDUs.


Despite the fact that frame preemption may well be applied on outgoing transmissions from a same node, its extension to incorporate transmissions from other APs/STAs would require a complex channel access mechanism. In that situation, and in presence of a time-sensitive traffic flow, it would be advisable to simply avoid the use of packet aggregation in the OBSS, even if that implied a severe throughput loss.

\begin{figure*}[]{}
    \centering
    \includegraphics[width=0.95\textwidth]{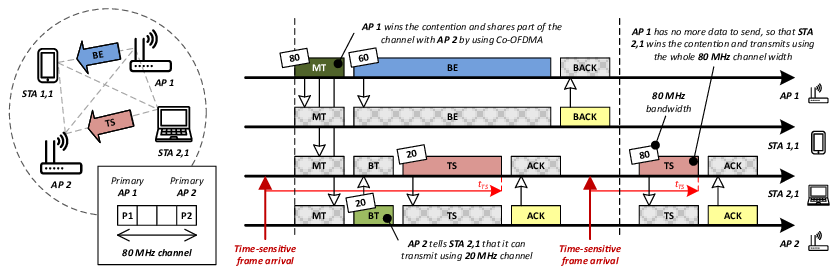}
    \caption{Multi-AP resource coordination based on Co-OFDMA, being \textit{MT}: Multi-AP trigger, \textit{BT}: Basic trigger, \textit{BE}: Best-effort frame, \textit{TS}: Time-sensitive frame, \textit{(B)ACK}: (Block) acknowledgement, and $t_{\text{TS}}$: Time-sensitive frame latency.}
    \label{fig:obss}
\end{figure*}

\subsection{Admission Control}

A traditional approach to protect time-sensitive traffic is to avoid channel overloading (i.e., to limit the traffic load, the number of traffic flows, and/or the number of STAs allowed to transmit data) in a given band and time period:

\begin{itemize}
    \item Whereas IEEE 802.11e admission control mechanisms limit the number of traffic flows per service class in a BSS, IEEE 802.11aa extends this capability to an entire OBSS. Both could be used by WiFi~7 alone or in combination with the traffic shaping solution previously introduced in subsection \textit{Traffic prioritization} to control how traffic arrives to the transmission buffer.
    \item Multi-link operation and the incorporation of the 6~GHz band in IEEE 802.11be foresee the emergence of traffic-aware multi-band admission control systems. For instance, depending on the network conditions and existing load, the 6 GHz band could be fully and exclusively dedicated to time-sensitive traffic.
\end{itemize}


Future admission control mechanisms may also be required to support scheduled operation. Thus, in the aforementioned example, the 6 GHz band would become even more exclusive, by only accepting time-sensitive traffic coming from devices able to operate in contention-free mode.

\subsection{Scheduled operation} 

Transmission of time-sensitive and non time-sensitive traffic could be performed on a periodic basis to isolate one from another. In this sense, two methods of scheduled access facilitating collision-free operation are already available in IEEE 802.11ax:

\begin{itemize}
    \item The trigger-based access allows the AP to schedule uplink MU transmissions. Future improvements could include a rate adaptation mechanism to increase the delivery ratio of time-sensitive frames as well as persistent allocation schemes to reduce control overhead caused by trigger signaling.
    \item By using the target wake time (TWT) mechanism, STAs adopt a \textit{wake time schedule} that makes them wake up on a periodic basis to transmit/receive data. 
\end{itemize}


Whereas the two aforementioned methods just determine the very moment in which the channel is accessed, the new MAC features fostered by TGbe could empower scheduled operation, especially if, as discussed before, only devices supporting those mechanisms are admitted in the 6 GHz band. In consequence, multi-link operation and OFDMA could play an important role by allocating devices' and network resources together with the computed schedule in function of the existing time-sensitive traffic load.

In brief, such a scheduled operation is key in terms of delay. However, the main obstacle that hinders its precise operation in the wireless domain continues to be the contention in the context of several OBSSs, which can only be effectively handled in combination with a proper multi-AP resource coordination strategy (as in Figure~\ref{fig:obss}, for instance).

\section{Use cases}
\label{usecases}

\begin{table}[]
\scriptsize
\centering
\caption{WiFi~7 low-latency use cases.} 
\label{tab:usecases}
\begin{tabular}{|l|c|c|c|}
\hline
\multicolumn{1}{|c|}{\multirow{2}{*}{\textbf{\begin{tabular}[c]{@{}c@{}}Sector and \\ use case\end{tabular}}}}    & \multicolumn{3}{c|}{\textbf{Requirements}}                                                                                                                                                                   \\ \cline{2-4} 
\multicolumn{1}{|c|}{}                                                                                            & \textbf{\begin{tabular}[c]{@{}c@{}}Latency\\ (ms)\end{tabular}} & \textbf{\begin{tabular}[c]{@{}c@{}}Reliability\\ (\%)\end{tabular}} & \textbf{\begin{tabular}[c]{@{}c@{}}Throughput\\ (Mbps)\end{tabular}} \\ \hline
{\textbf{MULTIMEDIA}}                                                                                        & \multicolumn{3}{l|}{}                                                                                                                                                                                        \\ \hline
\begin{tabular}[c]{@{}l@{}}Real-time high-quality\\ video streaming\end{tabular}                                  & 3 - 10                                                          & \textgreater 99.9                                                   & 5 - 25                                                               \\ \hline
Virtual Reality                                                                                              & 10 - 20                                                         & \textgreater 99.9                                                   & 25 - 500                                                             \\ \hline
Augmented Reality                                                                                             & 1 - 50                                                          & \textgreater 99.99                                                  & 1 - 200                                                              \\ \hline
Real-time pro graming                                                                                             & 5 - 50                                                          & \textgreater 99.9                                                   & \textgreater 3                                                       \\ \hline
Cloud gaming                                                                                                      & 5 - 50                                                          & \textgreater{}99.9                                                  & 10 - 35                                                              \\ \hline
{\textbf{HEALTH CARE}}                                                                                        & \multicolumn{3}{l|}{}                                                                                                                                                                                        \\ \hline
\begin{tabular}[c]{@{}l@{}}Telediagnosis, telemonitoring,\\ and telerehabilitation\end{tabular}                   & 50 - 200                                                        & \textgreater 99.9                                                   & 0.5 - 5                                                              \\ \hline
Telesurgery                                                                                                       & 1 - 10                                                          & \textgreater 99.9999                                                & $\sim$10                                                             \\ \hline
Exoskeletons and prosthetic hands                                                                                 & 5 - 20                                                          & \textgreater 99.999                                                 & 0.2 - 1                                                              \\ \hline
{\textbf{INDUSTRIAL}}                                                                                         & \multicolumn{3}{l|}{}                                                                                                                                                                                        \\ \hline
Process automation                                                                                                & 1 - 50                                                          & \textgreater 99.99                                                  & 0.1 - 5                                                              \\ \hline
Human machine interface                                                                                     & 50 - 200                                                        & \textgreater 99.9                                                   & $\sim$1                                                              \\ \hline
Tactile / Haptic technology                                                                                       & 1 - 5                                                           & \textgreater 99.999                                                 & $\sim$1                                                              \\ \hline
{\textbf{TRANSPORT}}                                                                                          & \multicolumn{3}{l|}{}                                                                                                                                                                                        \\ \hline
Real-time traffic information                                                                                     & 40 - 500                                                        & \textgreater 99                                                     & 0.1 - 1                                                              \\ \hline
\begin{tabular}[c]{@{}l@{}}Autonomous vehicle,\\ automated guided vehicle,\\ and drone control\end{tabular} & 10 - 100                                                        & \textgreater 99.9999                                                & 1 - 5                                                                \\ \hline
\begin{tabular}[c]{@{}l@{}}Remote-controlled vehicle \\ with video\end{tabular}                                   & 10 - 100                                                        & \textgreater 99.99                                                  & $\sim$10                                                             \\ \hline
\end{tabular}
\end{table}

The ability of WiFi~7 to support low-latency operation would open the door to multiple use cases. This section groups them into a set of productive sectors, details their performance requirements in Table~\ref{tab:usecases}, and discusses the suitability of using WiFi~7 with respect to other alternatives.

\subsection{Multimedia}

WiFi is nowadays the predominant Internet access technology for mobile devices in home and office environments running multimedia applications. The short-term evolution of this sector foresees the consolidation of more advanced time-sensitive services such as real-time high-quality 4K/8K audio and video streaming, virtual reality, augmented reality, cloud gaming, and interactive applications which will not be only targeted for entertainment, but also for educational and instructive purposes.

The already generalized adoption of WiFi technology indoors, its backward compatibility, and its distinctive features with respect to wired alternatives (that is, essentially, flexibility, simplicity, and mobility) suggest that the emergence of a low-latency operation mode for WiFi~7 would position it as a preferential option for upcoming multimedia use cases together with 5G enhanced Mobile Broadband.




\subsection{Health care}



Latest IT advances such as ultra high video resolution, Big Data and artificial intelligence will take health care to a next level, enabling a plethora of innovative applications in remote diagnosis (telediagnosis), treatment (telesurgery), and recovery (telemonitoring, telerehabilitation, exoskeletons, and prosthetic hands) for a wide set of diseases. WiFi~7 and 5G will play here again an important role as enablers of novel medical \textit{wearables} and devices intended for use in smart health care and home environments.

As for the specific use cases, some of them share common characteristics with the multimedia sector (e.g., telediagnosis, telemonitoring, and telerehabilitation). Due to their criticality, some others additionally impose extremely stringent network requirements in terms of end-to-end reliability, latency, and security (e.g., telesurgery). And lastly, a third group involving remote motion control with relatively low traffic load requires a fully deterministic approach (e.g., exoskeletons and prosthetic hands).

\subsection{Industrial}

During the next years, wireless networks will have increasing weight in the industry, leading a trend towards more flexible production sites and consolidating the Industry~4.0 concept. \textit{Connected factories} will then become a reality, involving monitoring, management, and direct control of machines, robots, and other industrial assets.

Future industrial communications will probably rely on the coexistence among wired (e.g., Fieldbus-based and Industrial Ethernet), wireless (from RFID to LoRa, to cite two examples), and 5G/6G-based cellular technologies. WiFi~7 is also expected to get a foothold in this sector, not only because of its inherited features (namely, flexibility, ease of installation, scalability, and interoperability), but also thanks to its new enhancements, particularly in terms of improved resource management and support to deterministic communications. 

\subsection{Transport}

Transport is experiencing such profound changes that future mobility will certainly be substantiated by automation, sustainability, road/air/sea safety, and energy efficiency. Real-time traffic information is starting to be served on a regular basis to drivers, using for instance city-wide WiFi deployments. Yet the upcoming revolution is being led by autonomous vehicles and automated guided vehicles, which will be able to transport people and goods thanks to their WiFi/5G connections without any human intervention. 

Next-generation vehicle communication and processing systems, such as vehicle-to-everything communication or advanced driver-assistance systems, will assist future transport systems on the basis of TSN and artificial intelligence. Hence, ensuring very high reliability and low latency in future transport applications will become crucial regardless the employed technology, due to the high relative speeds among end devices, and the continuous dynamism and low predictability of the outdoor environment.




\section{Conclusions}
\label{conclusions}

A new world of technological possibilities could make its way in a very varied range of sectors thanks to the integration of TSN and the well-established IEEE 802.11 technology. The most solid and promising exponent of this trend is IEEE 802.11be, actual precursor of future WiFi~7, which should be accompanied with a  well-defined and backward compatible  time-sensitive operation mode to support low-latency communications.

Although WiFi will never be able to guarantee fully deterministic communications because of its operation in license-exempt bands, there is still room to reduce the impact of all \textit{manageable} causes, both internal and external, that may increase latency. On the one hand, contention with external networks may be minimized by considering dynamic spectrum access such as non-contiguous channel bonding and multi-link operation, as well as cooperative AP strategies. On the other hand, prioritization and scheduling mechanisms inside the same WLAN may provide an effective solution to reduce the latency of time-sensitive traffic in the presence of large packets from best-effort flows.






\section*{Acknowledgment}

This work received funding from the Spanish government under the projects PGC2018-099959-B-100 and TEC2016-79510-P, and from the Catalan government through projects SGR-2017-1188 and SGR-2017-1739.


\bibliography{Bib}
\bibliographystyle{ieeetr}

\section*{Biographies}

\textbf{Toni Adame} is a Senior Researcher and Associate Lecturer in the Department of Information and Communication Technologies at the Universitat Pompeu Fabra. 

\textbf{Marc Carrascosa} is currently a PhD student in the Wireless Networking research group of the Department of Information and Communication Technologies at the Universitat Pompeu Fabra.

\textbf{Boris Bellalta} is an Associate Professor in the Department of Information and Communication Technologies (DTIC) at the Universitat Pompeu Fabra (UPF). He is the head of the Wireless Networking research group at DTIC/UPF.

\end{document}